\documentclass[amsmath,amssymb,aps,prl,reprint,superscriptaddress,nofootinbib,preprintnumbers]{revtex4-2}

\usepackage{amsmath,amssymb,amsthm,amsfonts}
\usepackage{graphicx,tabularx}
\usepackage{color}
\usepackage{multirow}
\usepackage{comment}
\usepackage{enumitem}
\usepackage{subfigure}
\usepackage{orcidlink}
\usepackage{cleveref}
\usepackage{slashed}
\usepackage[normalem]{ulem}
\usepackage{scalerel}
\usepackage{wrapfig}
\usepackage{cancel}
\usepackage{xcolor}
\usepackage{multirow}
\usepackage{setspace}
\usepackage{xspace}
\usepackage{lineno}

\newcommand{\PRE}[1]{{#1}} 

\newcommand{\be}{\begin{equation} \begin{aligned}}
\newcommand{\ee}{\end{aligned} \end{equation}}
\newcommand{\beqa}{\begin{eqnarray}}
\newcommand{\eeqa}{\end{eqnarray}}

\def\figureautorefname~#1\null{Fig.\,#1\null}
\def\tableautorefname~#1\null{Tab.\,#1\null}
\def\equationautorefname~#1\null{Eq.\,(#1)\null}

\newcommand{\quirk}{\ensuremath{\mathcal{Q}}\xspace}

\newcommand{\mquirk}{\ensuremath{m_\mathcal{Q}}\xspace}

\newcommand{\pt}{\ensuremath{p_\text{T}}\xspace}
\newcommand{\tz}{\ensuremath{t_0}\xspace}
\newcommand{\bst}{\ensuremath{\beta}\xspace}
\newcommand{\effsig}{\ensuremath{\sigma^\text{eff}_\text{fid}}\xspace}
\newcommand{\effsigPred}{\ensuremath{\sigma^\text{pred}_\text{fid}}\xspace}
\newcommand{\zkink}{\ensuremath{z_{\textrm{kink}}}\xspace}

\newcommand{\MHz}{\ensuremath{\,\text{MHz}}\xspace}

\newcommand{\ev}{\ensuremath{\,\text{eV}}\xspace}
\newcommand{\kev}{\ensuremath{\,\text{keV}}\xspace}

\newcommand{\gev}{\ensuremath{\,\text{GeV}}\xspace}
\newcommand{\tev}{\ensuremath{\,\text{TeV}}\xspace}

\newcommand{\ifb}{\ensuremath{\,\text{fb}^{-1}}\xspace}

\newcommand{\mm}{\ensuremath{\,\text{mm}}\xspace}
\newcommand{\cm}{\ensuremath{\,\text{cm}}\xspace}
\newcommand{\m}{\ensuremath{\,\text{m}}\xspace}

\newcommand{\ns}{\ensuremath{\,\text{ns}}\xspace}

\newcommand{\mrad}{\ensuremath{\,\text{mrad}}\xspace}

\RequirePackage[normalem]{ulem}

\crefname{section}{Sec.}{Secs.}
\crefname{figure}{Fig.}{Figs.}
\crefname{equation}{Eq.}{Eqs.}
\crefname{table}{Table}{Tables}
\crefname{appendix}{Appendix}{Appendices}

\begin{document}

\preprint{CERN-EP-2026-215}

\title{First Search for Quirks at the LHC with FASER}
\author{Roshan Mammen Abraham\,\orcidlink{0000-0003-4678-3808}}
\affiliation{Department of Physics and Astronomy, University of California, Irvine, CA 92697-4575, USA}

\author{Xiaocong Ai\,\orcidlink{0000-0003-3856-2415}}
\affiliation{School of Physics, Zhengzhou University, Zhengzhou 450001, China}
  
\author{Saul Alonso Monsalve\,\orcidlink{0000-0002-9678-7121}}
\affiliation{Institute for Particle Physics, ETH Z\"urich, Z\"urich 8093, Switzerland}

\author{John Anders\,\orcidlink{0000-0002-1846-0262}}
\affiliation{University of Liverpool, Liverpool L69 3BX, United Kingdom}

\author{Emma Kate Anderson\,\orcidlink{0000-0002-0161-4560}}
\affiliation{CERN, CH-1211 Geneva 23, Switzerland}

\author{Akitaka Ariga\,\orcidlink{0000-0002-6832-2466}}
\affiliation{Albert Einstein Center for Fundamental Physics, Laboratory for High Energy Physics, University of Bern, Sidlerstrasse 5, CH-3012 Bern, Switzerland}
\affiliation{Department of Physics, Chiba University, 1-33 Yayoi-cho Inage-ku, 263-8522 Chiba, Japan}

\author{Tomoko Ariga\,\orcidlink{0000-0001-9880-3562}}
\affiliation{Kyushu University, 744 Motooka, Nishi-ku, 819-0395 Fukuoka, Japan}

\author{Jeremy Atkinson\,\orcidlink{0009-0003-3287-2196}}
\affiliation{Albert Einstein Center for Fundamental Physics, Laboratory for High Energy Physics, University of Bern, Sidlerstrasse 5, CH-3012 Bern, Switzerland}

\author{Florian~U.~Bernlochner\,\orcidlink{0000-0001-8153-2719}}
\affiliation{Universit\"at Bonn, Regina-Pacis-Weg 3, D-53113 Bonn, Germany}

\author{Jianming Bian\,\orcidlink{0000-0003-3739-5424}}
\affiliation{Department of Physics and Astronomy, University of California, Irvine, CA 92697-4575, USA}

% ...

\collaboration{FASER Collaboration}

\begin{abstract}
\begin{center}\textbf{Abstract}\end{center}

We present the first LHC search for quirks, particles with Standard Model and QCD-like infracolor (IC) interactions. The analysis uses data collected in 2022--2024 by FASER, with a $186\ifb$ integrated luminosity of proton-proton collisions at a center-of-mass energy $\sqrt{s} = 13.6\tev$. Scintillator charge and timing information are used to search for slow pairs of quirks with no QCD color and electric charge $\pm 1$. No events are seen in this nearly background-free search, resulting in the first exclusion of quirks with masses above the weak scale for IC confinement scales in the broad range of $300 \ev<\Lambda<100\kev$.

\end{abstract}

\maketitle

\onecolumngrid
\begin{center}
\copyright~2026 CERN for the benefit of the FASER Collaboration. Reproduction of this article or parts of it is allowed as specified in the CC-BY-4.0 license.
\end{center}
\twocolumngrid

The Standard Model (SM) of particle physics is the current framework to describe the fundamental building blocks of the universe and their interactions. However, it faces many theoretical and experimental challenges, including its failure to resolve the gauge hierarchy problem or provide a viable particle candidate for cold dark matter.
A generic extension of the SM can be made by adding a new gauge group. While the Abelian gauge groups have been thoroughly investigated, non-Abelian groups are much less studied, despite being more generic. Such hidden non-Abelian sectors are further motivated by neutral naturalness models~\cite{Chacko:2005pe, Burdman:2006tz, Batell:2022tif}, where non-colored top partners, charged under a hidden gauge group, can ameliorate the gauge hierarchy problem.

Quirks~\cite{Kang:2008ea} are hypothetical matter particles charged under infracolor (IC), a new, QCD-like gauge group. IC models and quirks are motivated by the facts that IC glueballs are viable dark matter candidates~\cite{Juknevich:2009ji,Asadi:2025btr}, and the framework arises in attempts to resolve the gauge hierarchy problem~\cite{Burdman:2006tz,Cai:2008au}. Despite this motivation, there have been few direct searches for quirks at colliders. The D\O\ Collaboration searched for quirks at the Tevatron~\cite{D0:2010kkd}, but there has not yet been any dedicated search for quirks at the Large Hadron Collider (LHC), although there have been a number of proposed search strategies for ATLAS and CMS~\cite{Burdman:2008ek, Farina:2017cts, Knapen:2017kly, Evans:2018jmd, Sha:2024hzq, Forsyth:2025wks, Curtin:2025ngf} and LHCb~\cite{CidVidal:2020okg}.

For this letter, following the benchmark model studied in Ref.~\cite{Feng:2024zgp}, we consider the IC gauge group $\mathrm{SU}(N_{\mathrm{IC}})$ with a single quirk flavor in the $(1, 1, -1, N_{\mathrm{IC}})$ representation of $\mathrm{SU}_\mathrm{C}(3) \times \mathrm{SU}_\mathrm{L}(2) \times \mathrm{U}_\mathrm{Y}(1) \times \mathrm{SU}(N_\mathrm{IC})$, corresponding to the same SM gauge charges as right-handed charged leptons, with no SM color and electric charge $-1$. We take the number of infracolors to be $N_\mathrm{IC}$  = 2.  This quirk model is, then, completely specified by the quirk mass $m_\quirk$ and the IC confinement scale $\Lambda$. Quirks with SM color are not considered here, given the complexity of including QCD hadron emission, but they can be probed with a similar analysis.

Quirk-antiquirk pairs are dominantly produced at the LHC through Drell-Yan processes mediated by a photon or Z boson. At leading order, the quirk pair is produced with no transverse momentum (\pt), and thus close to the beam collision axis. For $\Lambda \ll \mquirk$, quirk pairs remain bound to each other via the IC force and oscillate around their common center of mass, while their high mass does not allow the creation of new quirk pairs out of the vacuum. The oscillation amplitude is approximately $1\cm\times(\mquirk/100\gev)(1\kev/\Lambda)^2$. The pair of quirks can travel long distances leaving exotic electromagnetic traces. Experimentally, quirk pairs can be identified as forward, long-lived signatures with trajectories different from those of SM particles. In particular, their weak-scale masses allow them to be produced with speeds significantly below the speed of light $c$. The IC force redirects part of each quirk's velocity into transverse oscillations, further reducing its longitudinal velocity. The resulting slow speed provides a powerful discriminant in the forward direction, as discussed in Ref.~\cite{Feng:2024zgp}.

The FASER detector~\cite{FASER:2022hcn} at the LHC is well suited to search for quirk signatures~\cite{Li:2021tsy,Feng:2024zgp}. It is located 480\m downstream of the ATLAS experiment's interaction point, aligned with the beam collision axis. The LHC dipole magnets and the material between ATLAS and FASER, including more than 100\m of rock, shield FASER from most SM particles. Muons and neutrinos are the only SM particles reaching FASER, making it a suitable environment to search for particles beyond the SM.

\begin{figure*}[htbp!]
\centering
\includegraphics[width=1.0\textwidth]{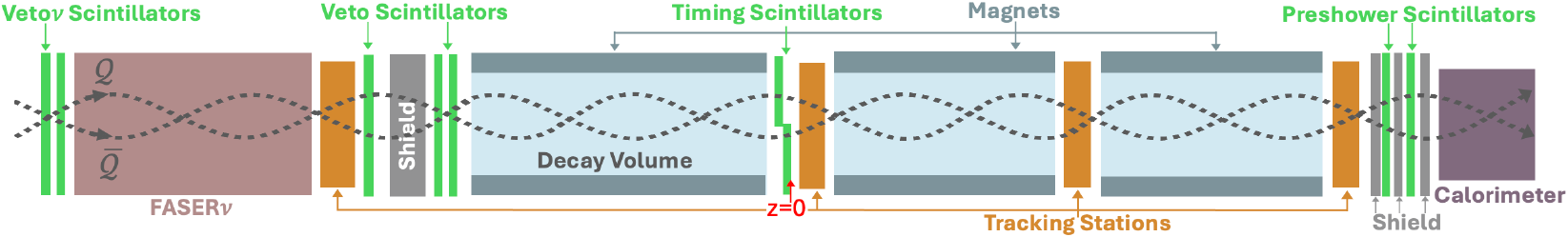}
\caption{Schematic of a quirk pair propagating through various components of the FASER detector. The detector geometry is not shown to scale.}
\label{fig:layout}
\end{figure*}

As shown in \cref{fig:layout}, the FASER detector consists of a 1.1-ton FASER$\nu$ emulsion detector, a decay volume, three permanent dipole magnets, a sampling electromagnetic calorimeter with lead absorbers, and four stations of a tracking spectrometer, each consisting of three $240\times240\,\text{mm}^2$ double-sided layers of silicon strip detectors which reconstruct track segments per station. The tracker readout window is limited to 75\ns. In addition, eight scintillators, placed across $5.78 ~\textrm{m}$ in the $z$-direction and each with a transverse size of at least $300\times300\,\text{mm}^2$ and thickness of 10 or 20\mm, provide information about both the charge and arrival time of charged particles, crucial in the detection of slow quirk pairs with charge deposition twice that of background muons. The scintillators and calorimeters are read out with photomultiplier tubes, and the corresponding waveforms are recorded for each event in a time window of at least $560\ns$, from which the arrival time and integrated charge can be reconstructed. The $z$-coordinate is defined along the ATLAS beam collision axis, with $z=0$ at the timing scintillator station. Transverse $x$ and $y$ coordinates are zero at the center of the detector.

This analysis is based on a data sample of proton-proton collisions at $\sqrt{s}=13.6\tev$ collected by FASER from 2022 to 2024, corresponding to an integrated luminosity of $186\ifb$. Events are triggered by measured charge above the noise level in any of the scintillators or the calorimeter. For each run, the scintillator time is calibrated to the average arrival time of muons, propagating at velocity $c$. The calibrated time of each scintillator is zero for a particle produced in a colliding bunch in ATLAS and traveling at speed $c$. A position-dependent correction, based on the tracker-inferred hit position, accounts for photon propagation within the scintillator. As detailed in \cref{app:timing-calibration}, this process leads to position- and charge-dependent timing resolutions per event and scintillator, with averages in the range of $0.25-0.6\ns$.

We simulate the Drell-Yan production of quirk pairs at ATLAS at leading order using \texttt{MadGraph5\_aMC@NLO}~\cite{Alwall:2014hca} interfaced with \texttt{Pythia}~8~\cite{Sjostrand:2007gs} for parton showering. Events within the fiducial phase space of $\theta<0.5\mrad$, where $\theta$ is the angle relative to the $z$-axis, are retained for quirk propagation. We use the LHCb tune~\cite{LHCb:2019qoc} of \texttt{Pythia 8} and apply a normalization correction derived from LHCb measurements of forward $Z$ boson production~\cite{LHCb:2021huf} to improve the modeling of initial-state radiation, to which the FASER signal acceptance is particularly sensitive.

Quirk propagation through material and LHC magnetic fields is simulated~\cite{eliwelch17_2026_21431221} following Ref.~\cite{Li:2019wce}. For $\Lambda>3\kev$, deflections from magnetic fields and ionization in material are negligible because the IC force dominates and rapid oscillations average out residual effects. The quirks are therefore propagated analytically without stepwise simulation, while ionization energy loss is still accounted for. Energy loss from IC glueball radiation and quirk-pair deflection outside the detector acceptance are included using survival probabilities parameterized by \mquirk, $\Lambda$, and the quirk-pair energy. The IC glueball-emission probability per oscillation is set to $\epsilon=0.1$, following Refs.~\cite{Feng:2024zgp,Evans:2018jmd}.

The interaction of quirks with the FASER detector is simulated in \texttt{Geant4}~\cite{GEANT4:2002zbu} using an adaptation of the quirks physics extension within the Athena software framework~\cite{atlas_collaboration_2021_4772550}, which accounts for the IC force in the propagation of individual quirks. The digitized detector response is reconstructed using the same pipeline as the collision data. Scintillator times are smeared according to the data-driven resolutions obtained after the timing calibration. Details of the quirk simulation in \texttt{Geant4} and the timing calibration are discussed in \cref{app:Geant4,app:timing-calibration}, respectively. 

The background model includes two sources. The first process, referred to as ``through-going muons", consists of muons coming from near the ATLAS interaction point and depositing charge in all FASER scintillators. The second, referred to as the ``kinked'' background, arises from muon-induced secondary showers in the detector material. Kinked events exhibit timing consistent with propagation at velocity $c$ upstream of a fitted position $z=z_{\textrm{kink}}$, followed by delayed signals in downstream scintillators due to the additional path length of the shower particles.
Contributions from other sources including cosmic muons and beam-induced backgrounds from the two LHC beams were found to be negligible. Simulated muon samples with kinematics predicted by FLUKA~\cite{Ferrari2005FLUKA,Battistoni2015FLUKA} are used for validations of the scintillator charge response, the modeling of timing, and the source of the kinked background.

Events are selected based on tracker activity, calorimeter energy, timing, and measured charge in the scintillators. Events with one track segment in each of the two upstream tracker stations and more than one segment in either of the two downstream stations are rejected to suppress the kinked background. The quirk oscillation pattern in the $x$-$y$ plane ranges from approximately linear to circular as the pair angular momentum increases. An ellipse is therefore fitted to the tracker-hit distribution. The reconstructed tracker-hit ellipse is required to have at least 95\% of its area contained within a tracker fiducial region defined by $|x|<115\mm$ and $|y|<115\mm$, inscribing a circle corresponding to polar angle $\theta\simeq0.24~\mathrm{mrad}$. Events for which the semi-minor axis of this ellipse is greater than 3\cm are rejected to suppress showers. 

To avoid calorimeter showers that could backsplash to the preshower scintillators, a loose upper cut of 15 (3)\gev on calorimeter activity for 2022--2023 (2024) events is applied, according to the noise level in this period with an integrated luminosity of $67\,(120)\ifb$.

The scintillator charge signal from a muon follows a Landau distribution convoluted with detector noise. A quirk pair typically produces approximately twice the measured charge of a single muon, allowing a minimum per-scintillator charge requirement to reject most through-going muons. Secondary showers, however, generally produce larger signals. A quirk-like charge (QLC) window is therefore optimized for each scintillator to maximize the expected signal significance. A preselection of data includes all signal selections mentioned so far, as well as requiring QLC in at least one Veto$\nu$ and one preshower scintillator. The single-muon QLC efficiency varies by scintillator and is weakly correlated across planes: 22.24\% of muons have QLC in the first Veto$\nu$ layer, 15.70\% in both Veto$\nu$ layers, and 0.01\% in all eight scintillators. Signal events must have QLC in all eight scintillators. This requirement is typically 99\% efficient for simulated quirks entering the tracker fiducial region.

\cref{fig:caltime-profile} shows the calibrated time as a function of scintillator position for a selected event in data compared to a simulated slow signal event. A linear regression extracts $\tz$, the arrival time at $z=0$ modulo the 25\ns LHC bunch spacing, as well as $\bst=v/c$, with $v$ being the extracted speed from the slope of the fitted line. Events with $\chi^2/N_\text{DoF}>2$ are rejected to ensure a precise extraction of \bst and \tz.

\begin{figure}[htbp!]
\centering
\includegraphics[width=0.47\textwidth]{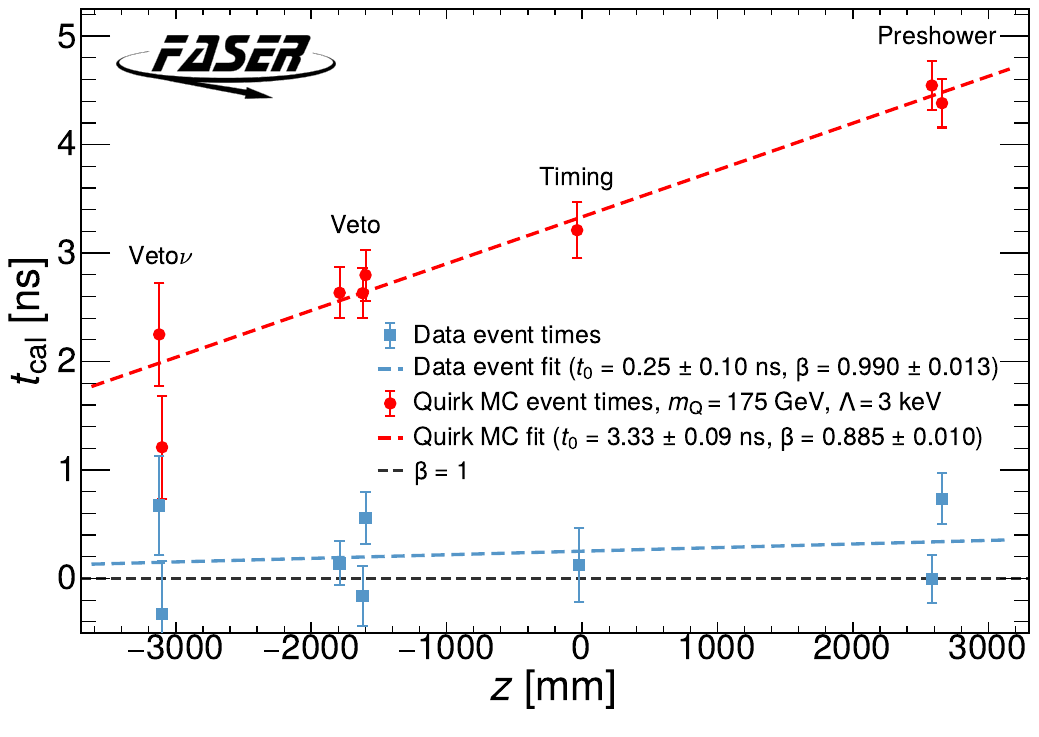}
\caption{Scintillator time vs. position for a selected event in data (blue) compared to a selected slow simulated quirk event (red). Scintillator times are calibrated to the arrival time of particles traveling at the speed of light (black).}
\label{fig:caltime-profile}
\end{figure}

\begin{figure*}[htbp!]
\centering
\subfigure[]{\includegraphics[width=0.491\textwidth]{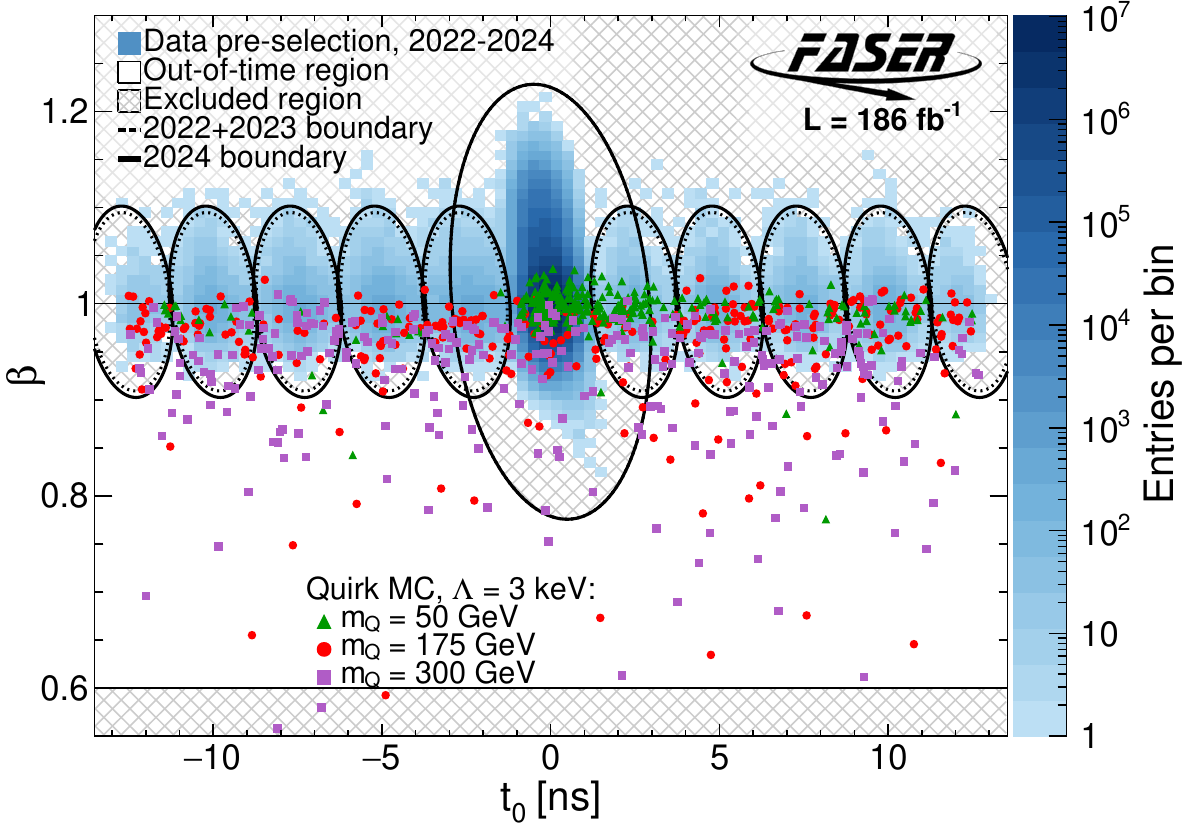}\label{fig:t0-speed}}
\hspace{0.005\textwidth}
\subfigure[]{\includegraphics[width=0.491\textwidth]{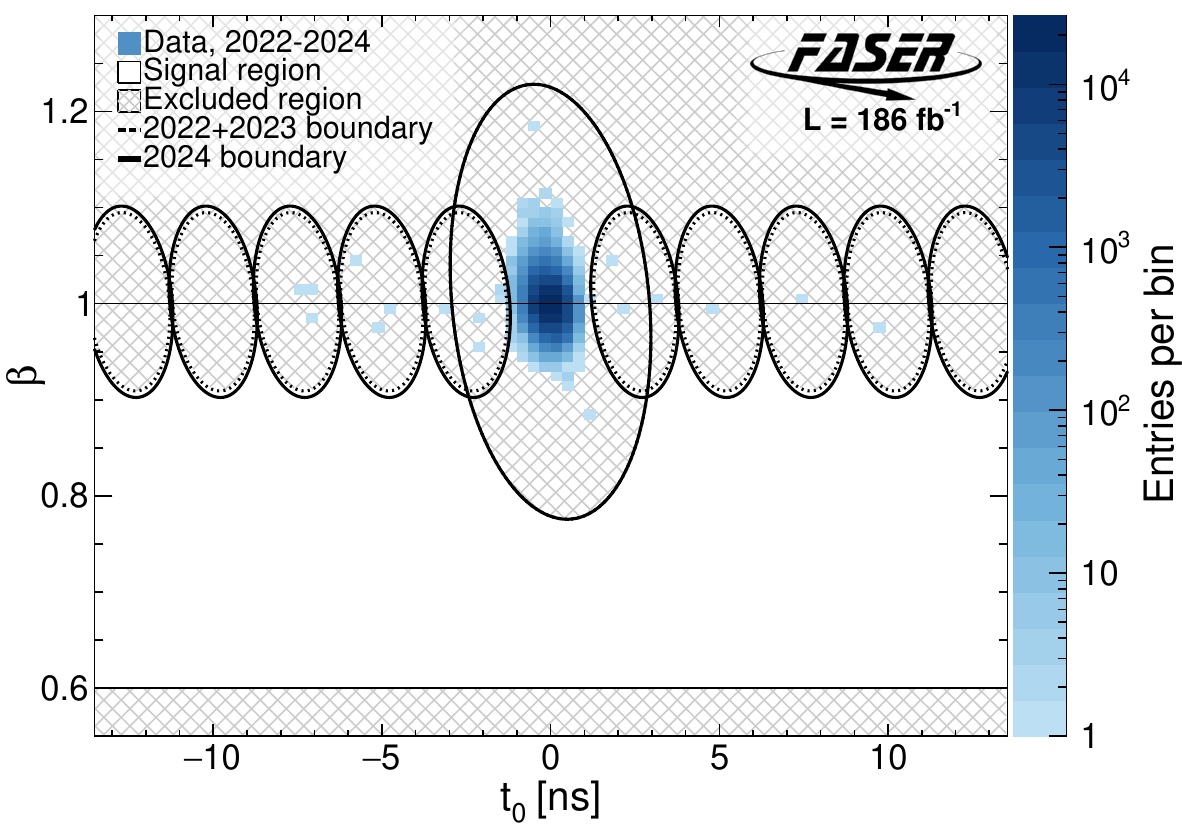}\label{fig:t0-speed-nScint8}}
\caption{(a) Preselected data (blue) and three representative signal models, each containing 300 events (green, red, purple) in the $(\tz, \beta)$ plane, where \tz is the arrival time modulo 25 ns, and $\beta = v/c$ is the reconstructed speed in units of $c$. (b) The same distribution for data after requiring QLC in all eight scintillators. The ellipses define the in-time regions for the two data-taking periods, with events above their lower halves classified as in-time and those below as out-of-time. The unhatched region in (b) is the signal region of this analysis.}
\label{fig:t0-speed-combined}
\end{figure*}

\cref{fig:t0-speed} shows the distribution of arrival times \tz and velocities \bst for preselected events in data, compared to a few representative quirk models, each with 300 simulated events. Because the per-run timing calibration is referenced to the mean arrival time, the muon-dominated data cluster near $\tz=0$ and $\bst=1$. Delayed radio-frequency (RF) satellite-bucket events~\cite{Jeff:1459107} arise when the 400\MHz RF cavities push a small fraction of protons from the main colliding bucket into the neighboring RF buckets, creating satellite bunches whose collisions occur at times offset by integer multiples of the 2.5\ns RF period relative to the nominal collisions. Massive quirks, however, are more likely to be slow and arrive out of time, possibly later than the 25\ns LHC bunch spacing.

Through-going muons are rejected using a combined arrival-time and speed selection, with ellipses in the $t_0$--$\beta$ plane defining background regions excluded from the final selection. Using speed alone removes substantial signal near the nominal bunch, while arrival time alone is susceptible to satellite-bunch backgrounds. The preselected data distribution is fitted with a two-dimensional Gaussian, whose $1\sigma$ ellipse is scaled by a factor $n$ chosen to maximize the signal significance against the background model. This gives $n=10.0$ for the main timing ellipse shown in \cref{fig:t0-speed}. Satellite events are rejected using 2.5\ns-spaced copies of the main ellipse, with their scaling factor optimized similarly, giving $n=4.1$ for 2022--2023 and $n=4.4$ for 2024. The optimization is performed separately for the two periods because differences in the LHC optics and collision crossing angle lead to differences in background-muon kinematics. Together, the ellipses define the excluded in-time region, with their lower halves forming its lower boundary.

Events are additionally required to have $0.6<\bst<1$. The lower bound allows sufficient propagation time through all tracking stations within their readout time window, while the upper bound removes a region predominantly populated by background events.

The final allowed, out-of-time region is shown unhatched in \cref{fig:t0-speed}. Depending on \mquirk, $0.03-0.04\%$ of signal events produced at the ATLAS interaction point reach FASER within the tracker fiducial region. The efficiency of reconstructing and selecting the remaining events is approximately 1\% (10\%) for $\mquirk=50\, (175)\gev$.

Statistical fluctuations in timing information and scintillator charge can lead to inefficiencies in rejection of through-going muons. This background is estimated with a data-driven ABCD method. Regions A--D span $0.6<\bst<1.0$: A and C lie outside the timing ellipses, while B and D lie inside; A and B require QLC in all scintillators, whereas C and D require QLC in 2--5 scintillators. The estimated background rate in the signal region A is fixed to $N_\text{A}=N_\text{B}N_\text{C}/N_\text{D}$, giving $0.046\pm 0.010$ in 2022--2023 and $0.082\pm 0.014$ in 2024. The uncertainties are determined by the Poisson uncertainties on the number of events in each control region.
This method relies on negligible correlation between charge and timing variables, which was confirmed by subdividing various control regions and repeating a similar method. Moreover, signal simulation studies confirm that there is negligible contamination of the signal in regions B--D.

The kinked-background rate is estimated using preselected data events with $n>5.5$ that populate the non-Gaussian tail within the main ellipse of \cref{fig:t0-speed-combined}. Their rate density falls approximately as a power law in $n$ and is extrapolated beyond the ellipse into the signal region. Kinked events have showers produced in both upstream and downstream detector material, but the downstream contribution is found to be negligible. The extrapolated rate is scaled by the probability that the upstream muon deposits QLC in the Veto$\nu$ scintillators, as described in \cref{app:kinked}. The kinked background rate in the signal region is found to be 0.033 (0.051) events, with a 68\% confidence level (CL) interval of [0.0015, 0.080] ([0.0036, 0.113]) in the 2022--2023 (2024) data-taking period.

The systematic uncertainties on the signal normalization depend on mass and confinement scale. A 2-4\% uncertainty is assigned to the inclusive production cross section, from QCD scales and PDF uncertainties. Initial- and final-state radiation during the production affect the \pt of the quirk pair and consequently the fiducial acceptance. Variations of shower scales in \texttt{Pythia}~8 by factors of 0.5 and 2 result in approximately 25\% uncertainty on the signal normalization. The uncertainty from IC glueball radiation is 1--20\%, relevant only for $\Lambda \gtrsim 30\kev$ and derived by modifying the radiation probability $\epsilon$ by $\pm30\%$. The uncertainty on the efficiency of requiring a good linear regression in the extraction of \tz and \bst is 1--5\%. The finite-simulation-sample systematic uncertainty is approximately 6\% for most relevant signal models, but up to 28\% for small confinement scales where the large oscillation amplitude translates to low signal acceptance. Finally, the uncertainty in the luminosity measurement is 2.2\%~\cite{ATLAS:2022hro,ATLAS:ATL-DAPR-PUB-2025-001}.

\cref{fig:t0-speed-nScint8} shows the remaining events in data after applying all analysis selections except timing requirements. The unhatched region represents the final signal region of this analysis containing zero observed events. We proceed to set limits on the effective fiducial cross section, \effsig, and signal strength, $\mu(\Lambda,\mquirk) = \effsig / \effsigPred$, using a fully frequentist ($\text{CL}_\text{s}$) method~\cite{Read:451614} within the CMS \textsc{Combine} framework~\cite{CMS:2024onh}. The effective fiducial cross section, \effsig, is defined as the cross section within the fiducial phase space of $\theta<0.5\mrad$ weighted by the theoretically calculated probability of survival from IC glueball radiation in a 480-meter propagation of the quirk pair towards FASER. It is theoretically predicted to be \effsigPred.

\begin{figure}[htbp!]
\centering
\includegraphics[width=0.49\textwidth]{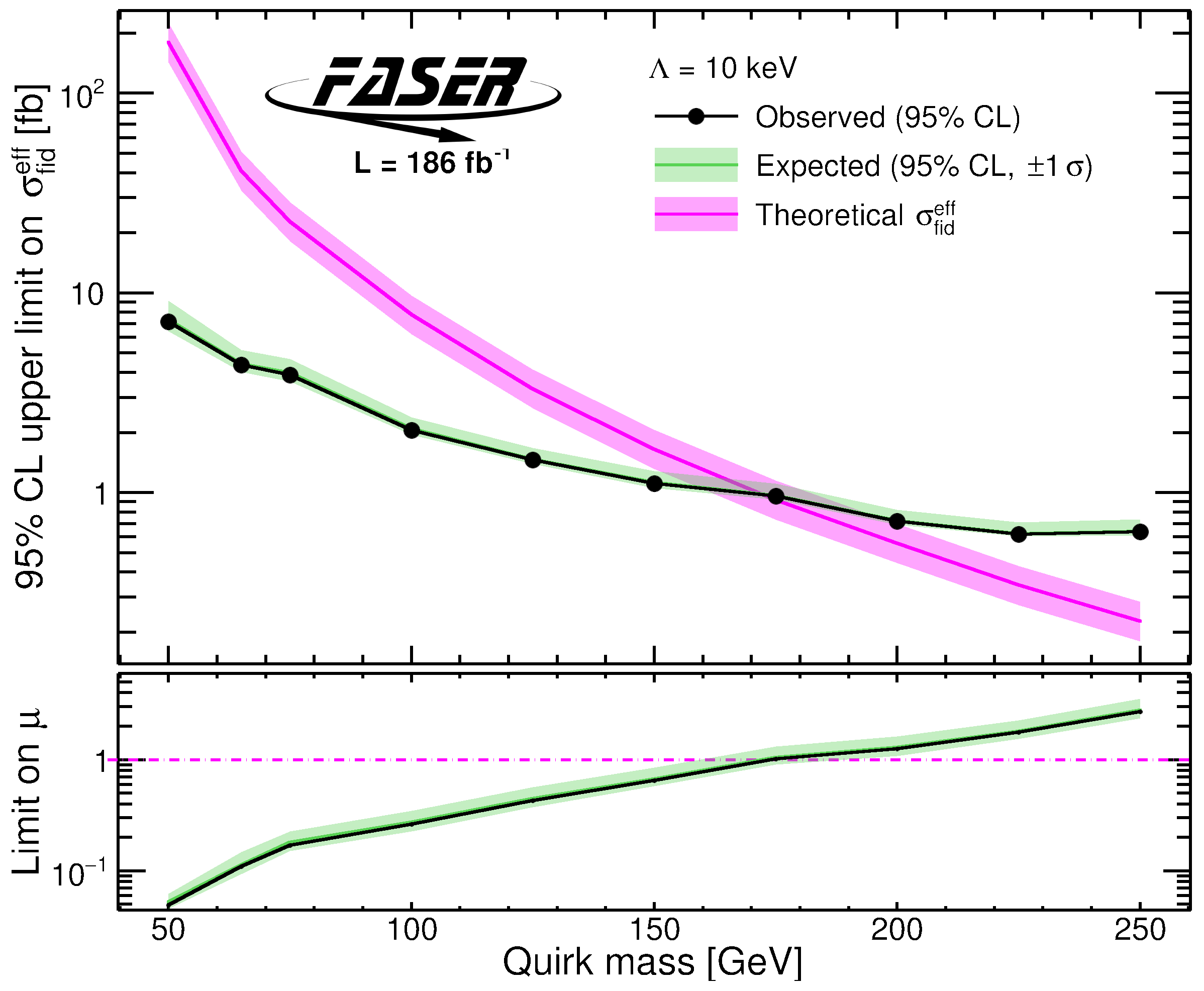}
\caption{Top: Observed (black) and expected (green) 95\% CL upper limit on the effective fiducial cross section, \effsig, compared to the theoretical prediction \effsigPred (magenta). Bottom: 95\% CL upper limit on $\mu(\Lambda,\mquirk) = \effsig / \effsigPred$, accounting for all theoretical and experimental uncertainties.}
\label{fig:1D-xsection}
\end{figure}

\cref{fig:1D-xsection} shows the observed and expected 95\% CL upper limits on \effsig as a function of the quirk mass for a confinement scale of 10\kev, compared to the theoretical prediction. It also shows the 95\% CL upper limit on $\mu(\Lambda,\mquirk)$, with the expected uncertainty band including all experimental and theoretical uncertainties. Masses up to 179\gev are excluded for $\Lambda=10\kev$.

\cref{fig:limits} shows the observed and expected excluded regions in the $\Lambda$, $\mquirk$ signal parameter space compared to previously excluded regions~\cite{CMS:2017kku,Farina:2017cts,D0:2010kkd}. For the first time, quirk models with masses beyond the electroweak scale are excluded for a wide range of confinement scales. Tabulated results, including limits on \effsig, are provided in the HEPData record for this analysis~\cite{HEPData}.

\begin{figure}[htbp!]
\centering
\includegraphics[width=0.49\textwidth]{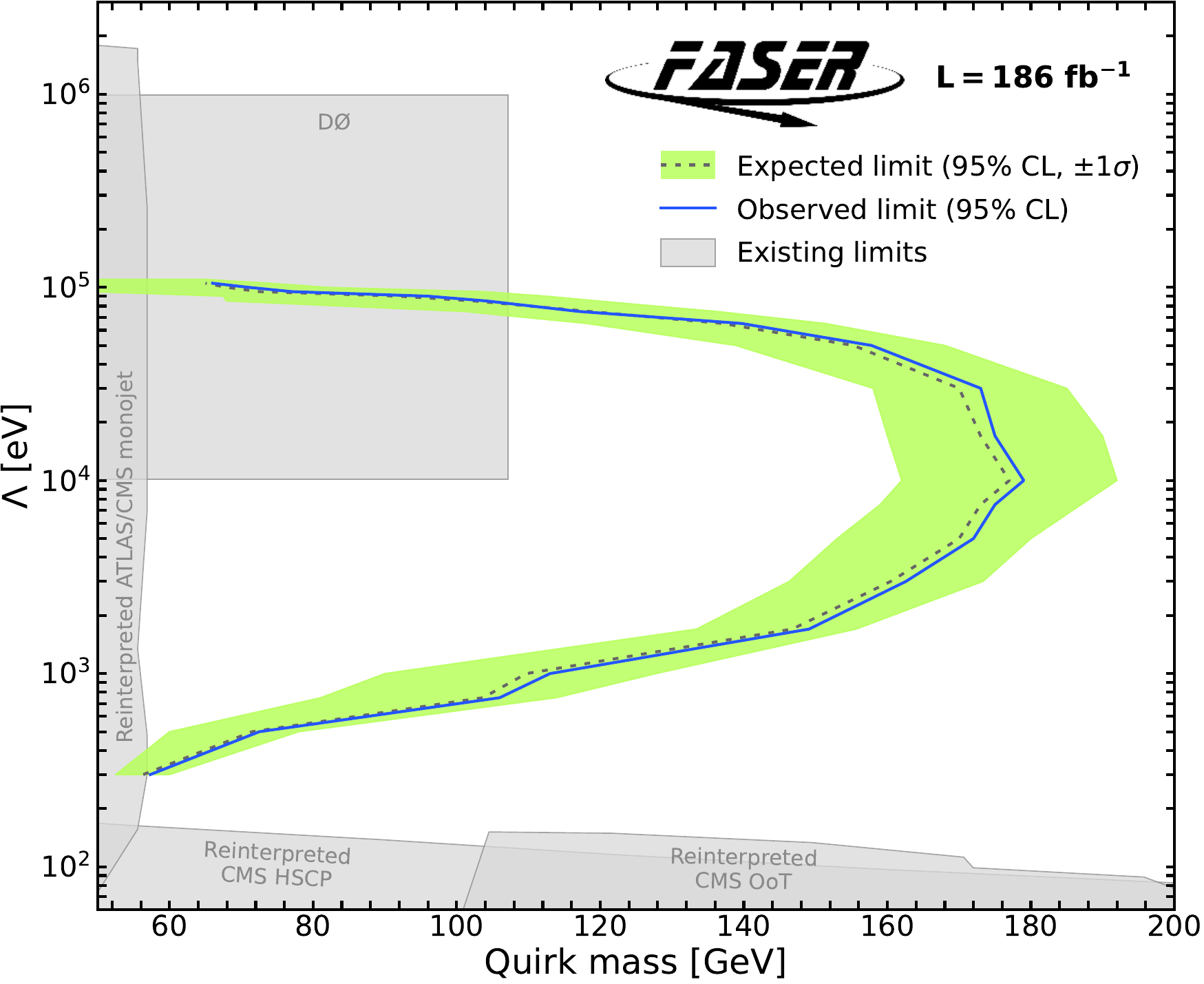}
\caption{Observed (solid blue) and expected (dashed blue) 95\% CL exclusion contours in the $(m_\quirk, \Lambda)$ plane, along with previously excluded regions (gray)~\cite{CMS:2017kku,Farina:2017cts,D0:2010kkd}. The green band reflects the $\pm1\sigma$ uncertainty on the expected limit.}
\label{fig:limits}
\end{figure}

In summary, a search for quirk particles at the LHC was performed using the data recorded by FASER in 2022--2024, approximately half of the recorded data in Run 3 of the LHC. As no quirk candidates were observed, lower limits were set on the quirk mass for a wide range of confinement scales covering three orders of magnitude. As the first direct search for quirks at the LHC, this analysis sets a precedent and offers a baseline methodology for similar searches in other collider experiments.

\section*{Acknowledgments}
\label{sec:Acknowledgments}

We thank CERN for the excellent performance of the LHC and the technical and administrative staff members at all FASER institutions for their contributions to the success of the FASER experiment. We thank the ATLAS Collaboration for providing us with accurate luminosity estimates for the Run 3 LHC $pp$ collision data. We thank the CERN STI group for providing detailed \texttt{FLUKA} simulations of the muon fluence at FASER. We also acknowledge the ATLAS Collaboration software, Athena, on which FASER’s offline software system is based. Finally, we thank James Black and Timothy Nelson for their work in developing the comprehensive \texttt{Geant4} quirks extension within the \texttt{Athena} software framework.

This work was supported in part by Heising-Simons Foundation Grant Nos.~2019-1179 and 2020-1840, Simons Foundation Grant No.~623683, JSPS KAKENHI Grant Nos.~19H01909, 22H01233, 20K23373, 23H00103, 20H01919, and 21H00082, the joint research program of the Institute of Materials and Systems for Sustainability, ERC Consolidator Grant No.~101002690, DFG grant~SCHO~1527/13-1, Royal Society Grant No.~URF$\backslash$R1$\backslash$201519, UK Science and Technology Funding Councils Grant No.~ST/ T505870/1, the National Natural Science Foundation of China, Tsinghua University Initiative Scientific Research Program, and the Swiss National Science Foundation.

\setcounter{secnumdepth}{1}
\appendix
\crefalias{section}{appendix}

\section{Simulation of Quirks in \texttt{Geant4}}
\label{app:Geant4}

Quirk propagation through the detector is simulated in \texttt{Geant4} using an adaptation of the \texttt{Simulation/G4Extensions/Quirks} physics extension from the \texttt{Athena} software framework ~\cite{atlas_collaboration_2021_4772550}. The quirk pair is treated as a coupled two-body system connected by an IC string. A custom \texttt{G4UserStackingAction} class, which controls the order in which tracks are processed during the \texttt{Geant4} event, assigns the quirk and antiquirk to separate waiting categories, enforcing alternating propagation. After each step, the string configuration is updated before the partner track is advanced, so, the pair is transported as a coupled system. The string force is applied by a dedicated transportation process, while standard \texttt{Geant4} electromagnetic processes describe material interactions.

The implementation is validated with three checks. First, track continuity is verified by comparing trajectories after upstream propagation with those at the detector simulation interface. Second, the scintillator charge response is compared with a straight-line reference using the same kinematics and relevant electromagnetic processes, confirming the expected response: similar to that of two minimum ionizing particles (MIPs). Third, the scintillator timing response is compared with the same reference. The charge and timing distributions are consistent within differences expected from the oscillatory quirk motion, which changes scintillator intersections and path lengths. These checks validate the detector-response quantities used in the analysis.

\section{Timing Calibration}
\label{app:timing-calibration}

Reconstructed scintillator times are measured relative to the LHC clock and are calibrated using high-energy, single-track muon events. Run-dependent reference offsets are first subtracted to remove relative timing shifts between channels and data-taking periods. The resulting times are mapped to a common 25\ns bunch period and the remaining bunch-index ambiguity is resolved using the absolute digitizer times. Position-dependent photon propagation times are measured in $2.5\times2.5~\mathrm{cm}^{2}$ bins across each scintillator, and the calibrated hit time is defined as $t_{\mathrm{cal}}=t-\overline{t}(x,y)$. 

For each scintillator plane, a \textit{leave-one-out} (LOO) estimate of the event arrival time at $z=0$, $t_{0}^{\mathrm{LOO}}$, is reconstructed from the measured times in all other valid planes, excluding the plane under study. The LOO procedure is iterated during the calibration, reducing the width of the residual distribution until it converges to the approximate intrinsic timing resolution. The timing uncertainty is then parametrized as a function of hit position and measured charge by fitting its charge dependence in each spatial bin to a power law.

The data-derived timing resolutions are used to smear the simulated hit times before applying the timing calibration. The tracker hits are used to define a containment ellipse for the expected hit region in each scintillator plane, over which the calibration values are averaged, with their spatial variation included as an additional uncertainty. This region is small and approximately circular for most muons and high-$\Lambda$ quirk events, while for low-$\Lambda$ quirks it ranges from a narrow, elongated ellipse to a large circle, depending on the angular momentum of the quirk system.

The resolutions of the calibrated time, $t_{\mathrm{cal}}$, and the uncalibrated raw time, $t_{\mathrm{raw}}$, are reported using muon data. For each scintillator plane, a LOO estimate of the arrival time at $z=0$, $t_{0,\mathrm{cal}}^{\mathrm{LOO}}$, is reconstructed from the calibrated times and uncertainties of all other valid planes, excluding the plane under study, assuming $v=c$. $t_{0,\mathrm{cal}}^{\mathrm{LOO}}$ is subtracted event by event from both $t_{\mathrm{cal}}$ and $t_{\mathrm{raw}}$ in the excluded plane to correct for the arrival-time due to the intrinsic collision-time spread of approximately $180~\mathrm{ps}$ arising from the finite bunch length. Gaussian fits are performed to the cores of the resulting time distributions, and the mean variance of the corresponding LOO arrival-time estimate, $\langle \sigma_{t_{0},\mathrm{LOO}}^{2}\rangle$, is subtracted in quadrature from the fitted widths. Using the calibrated event estimate for both measurements places them on the same reference, although it slightly improves the apparent raw time resolution relative to a fully independent raw time estimate.

The resulting typical scintillator timing resolutions are shown in \cref{fig:timing_resolutions}. The timing resolution is slightly worse for Veto$\nu$ scintillators as the light is extracted using a wavelength-shifting rod which widens the timing spread. The widths of the reconstructed $t_0$ and $\beta$ distributions are shown on the right-hand side of \cref{fig:timing_resolutions}, with the $t_0$ distribution including the intrinsic collision-time spread. For quirk signals, the scintillator charge is typically about two MIPs, corresponding to smaller charge-parameterized timing uncertainties than for the muon sample. 

\begin{figure}[htbp!]
    \centering
    \includegraphics[width=1.0\linewidth]{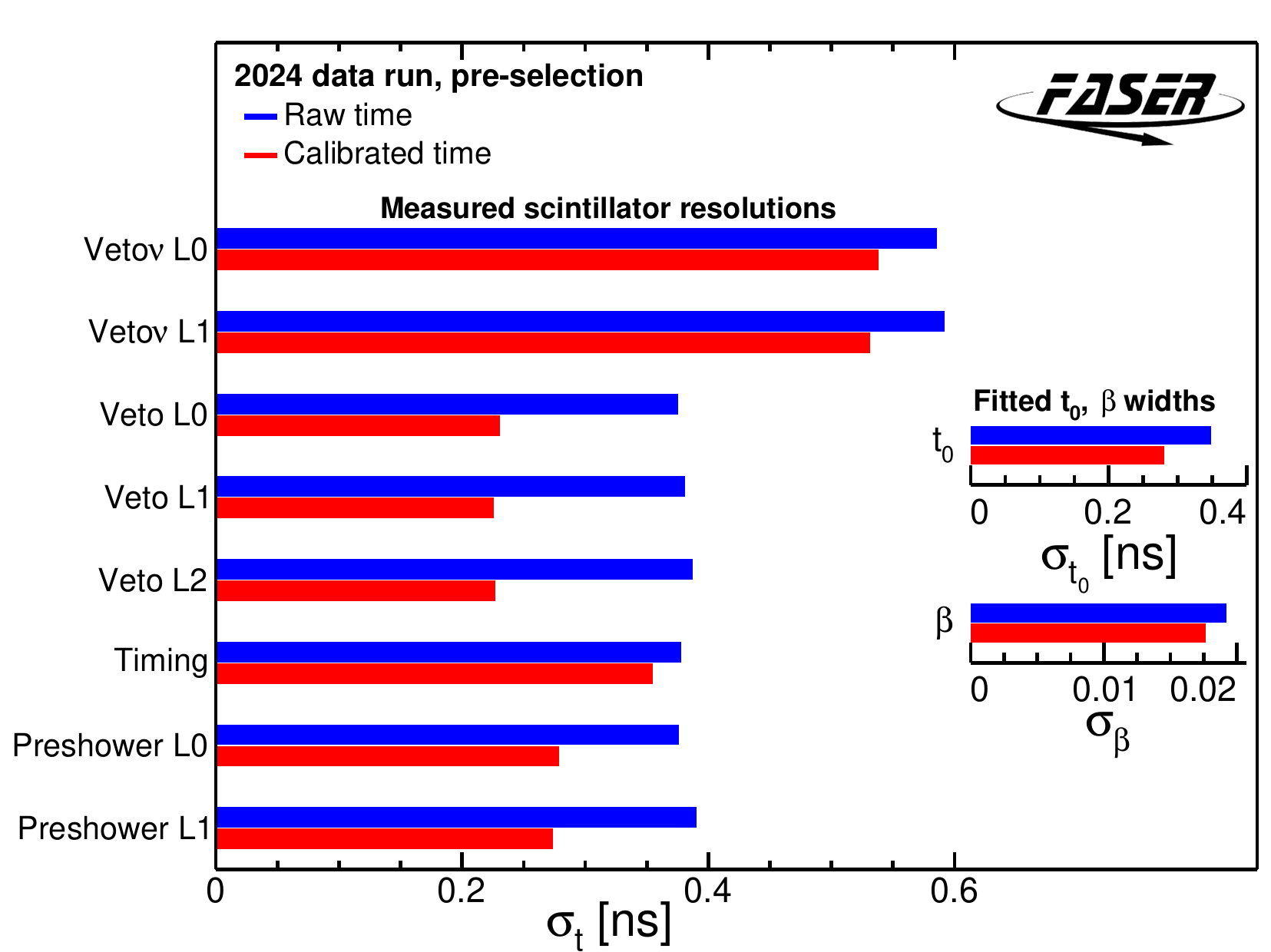}
    \caption{Measured timing resolutions for a 2024 data run. The right panels show the widths of the fitted $t_{0}$ and $\beta$ distributions for the raw and calibrated timing inputs. The $t_0$ distributions include collision-time spread.}
    \label{fig:timing_resolutions}
\end{figure}

\section{Kinked Background}
\label{app:kinked}

The kinked background arises when a muon traversing FASER at $v=c$ interacts in the detector and produces secondary shower particles. The additional path length accumulated by the shower particles delays the downstream scintillator signals, changing the timing profile from the $v=c$ behavior of the incoming muon to an apparently slower downstream propagation. Events with this characteristic kink can mimic the timing signature of a slow particle and appear at large $n$ relative to the nominal through-going-muon distribution in the $t_0$--$\beta$ plane.

Muon simulation identifies two main sources of such events: showers initiated in the FASER$\nu$ tungsten and hadronic backsplash from the calorimeter lead. To distinguish these sources, the scintillator times are fit with a piecewise timing model with three free parameters: the arrival time, the kink $z$-position \zkink, and the apparent downstream speed, with $v=c$ fixed upstream of \zkink. An example fit is shown in~\cref{fig:kinked_timing_ex}.

\begin{figure}[htbp!]
\centering
\includegraphics[width=0.49\textwidth]{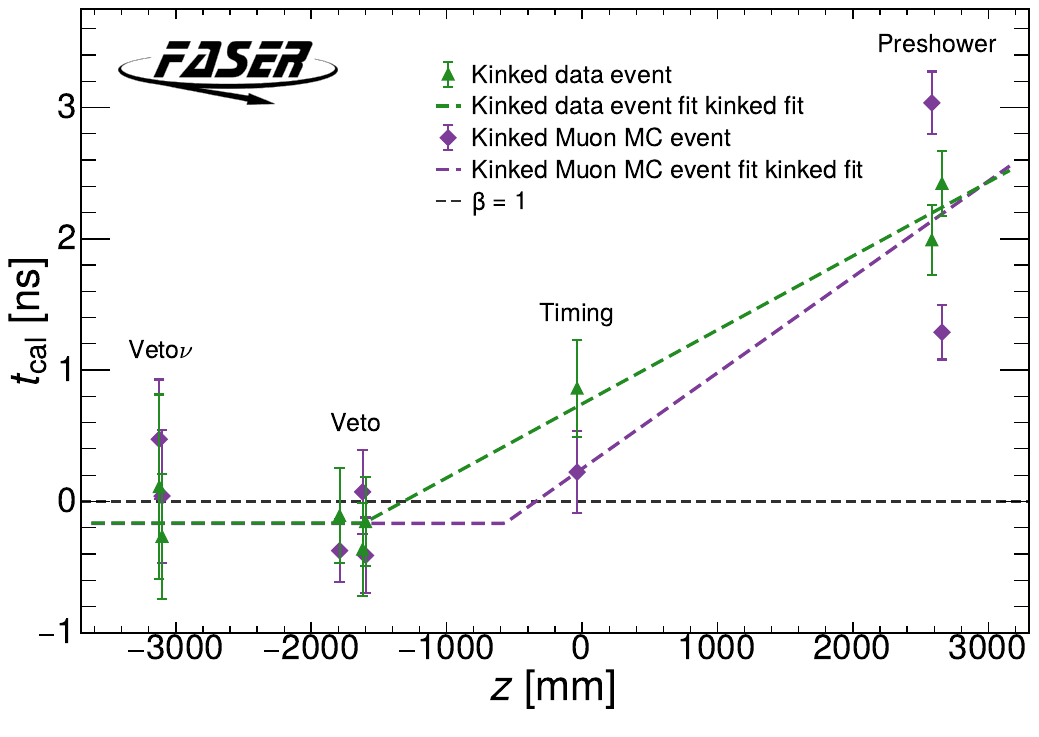}
\caption{Calibrated scintillator times for a candidate kinked-background event in data, compared to a simulated event in the same category. The kinked timing fit is used to extract the effective shower $z$-position \zkink.}
\label{fig:kinked_timing_ex}
\end{figure}

The fitted kink $z$-position \zkink is used to classify the likely shower origin, with $\zkink<0$ assigned to the upstream FASER$\nu$ category and $\zkink>0$ to the downstream calorimeter category.

To transfer the kinked background rate to the signal region, scintillators downstream of \zkink are conservatively assumed to satisfy the QLC requirement, while scintillators upstream of \zkink are assigned a probability of producing QLC signatures measured from data, $P(\textrm{QLC})$. This makes the upstream category the more pessimistic component estimate, since here $P(\textrm{QLC})$ = $P(\textrm{QLC in Veto}\nu)=0.157$ and all downstream scintillators are assumed to satisfy the QLC requirement. 

The kinked-background rate in the out-of-time region is estimated from data before applying the final timing and charge requirements. In both data and muon simulation, the kinked-event density falls approximately as a power law in $n$, motivating a power-law model for the extrapolation. The density in the preselected sample is fit for $n>5.5$, where contamination from through-going muons is suppressed; including lower-$n$ events would bias the fitted slope and underestimate the background at large $n$. No events in this fit region are classified as downstream, largely due to the calorimeter energy requirement, and the corresponding transfer factor would further suppress this contribution; only the upstream category is therefore considered in the background estimate.

The 2024 upstream kinked-background distribution and corresponding fit are shown in \cref{fig:kinked_upstream_2024}. The fitted density is integrated over the nominal out-of-time signal-region range in $n$. The same fitted density is integrated over the corresponding satellite out-of-time regions, with the satellite integral scaled by the measured ratio of events in the nominal bunch period to those in the satellite bunch periods. The sum of the nominal and scaled satellite integrals gives the upstream kinked yield before the final QLC requirement. This yield is then multiplied by the probability for QLC in the Veto$\nu$ scintillators.

\begin{figure}[htbp!]
    \centering
    \includegraphics[width=1.0\linewidth]{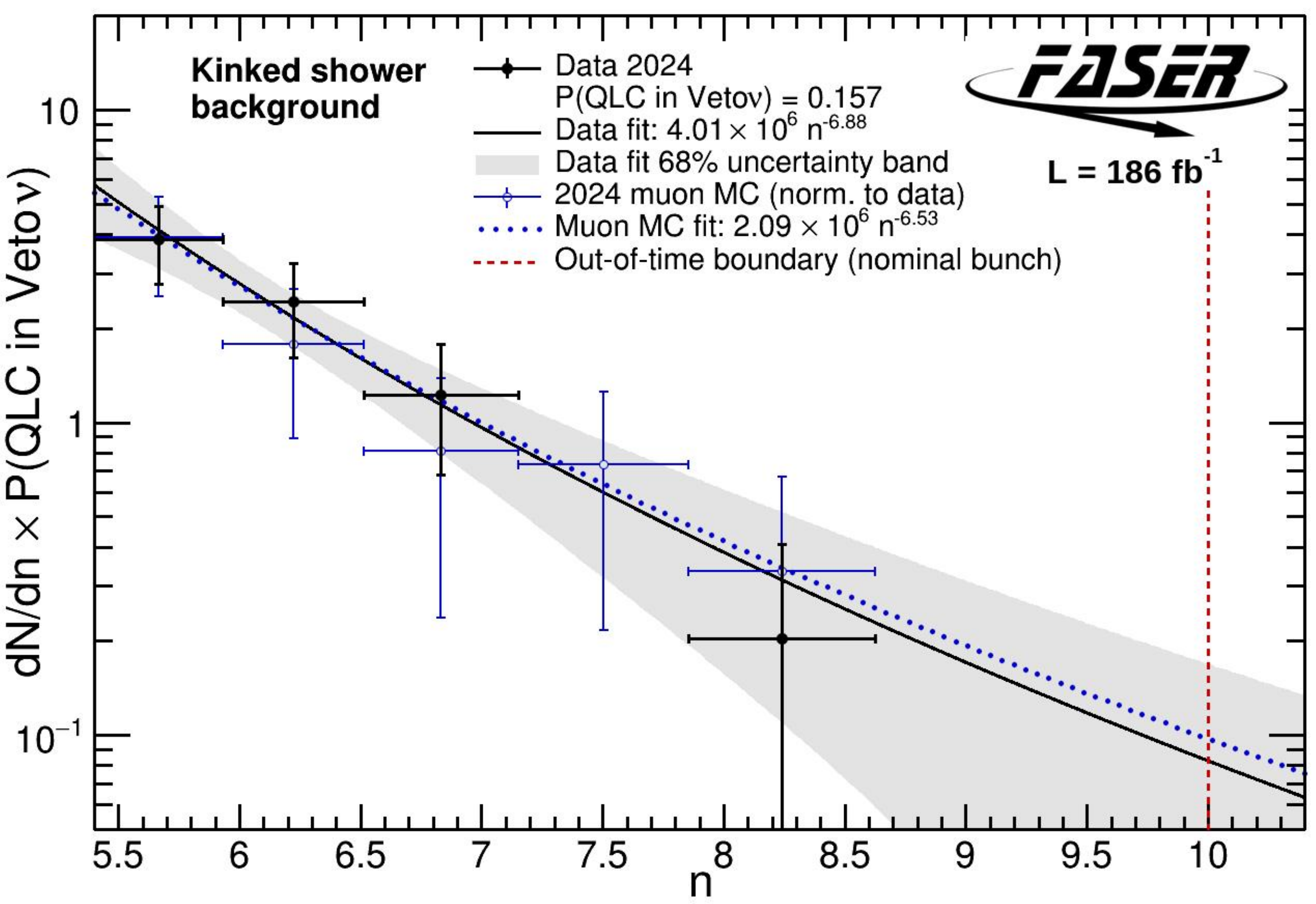}
    \caption{The 2024 upstream kinked-background estimate. The black points show the data event density multiplied by the Veto$\nu$ QLC probability, $P(\textrm{QLC in Veto}\nu)=0.157$, as a function of the ellipse-scaling variable $n$. The solid black curve shows the power-law fit used to extrapolate the kinked background into the signal region, and the gray band shows the 68\% uncertainty from the fit. The blue points show the 2024 muon simulation normalized to the data for a shape comparison, with the dotted blue curve showing the corresponding power-law fit. The dashed red line indicates the nominal out-of-time signal-region boundary.}
    \label{fig:kinked_upstream_2024}
\end{figure}

\bibliography{references}

\end{document}